\documentclass[12pt]{iopart}

\usepackage{graphicx}
\usepackage{verbatim}
\usepackage{color}
\usepackage{subfigure} 
\usepackage{dcolumn}
\usepackage{bm}
\usepackage[dvips]{epsfig}
\usepackage{bbm}

\begin{document}

\title[Collective  dynamics in crystalline polymorphs of ZnCl$_{2}$...]{Collective  dynamics in crystalline polymorphs of ZnCl$_{2}$: potential modelling and inelastic neutron scattering study}

\author{A Sen\footnote[1]{Present address: Department of Physics, Indian Institute of Technology, Kanpur 208016, India} 
\footnote[2]{Email: asen@iitk.ac.in}, Mala N. Rao, R Mittal and S. L. Chaplot\footnote[3]{Email: chaplot@magnum.barc.ernet.in}} 

\address{Solid State Physics Division, Bhabha Atomic Research Centre,
Trombay, Mumbai 400085, India}

\begin{abstract}
We  report a phonon density of states measurement of $\alpha$-ZnCl$_{2}$
using the coherent inelastic neutron scattering technique and a  lattice
dynamical  calculation  in four crystalline phases of ZnCl$_{2}$ using a
transferable  interatomic  potential.   The  model  calculations  agree
reasonably  well with  the  available  experimental data on the structures,
specific heat, Raman frequencies and their pressure variation in various
crystalline phases. The calculated results have been able to provide  a
fair description  of  the vibrational as well as the thermodynamic
properties of ZnCl$_{2}$ in all its four phases.
\end{abstract}


\section{Introduction}
Among  the  metal  halides  of  form  MX$_{2}$  (M:  metal,  X: halide),
ZnCl$_{2}$ continues to draw attention in current research  for  several
reasons.  Firstly,  the  crystalline  ZnCl$_{2}$  has  four structurally
determined polymorphs\cite{breh,bryn}  at  the  ambient  pressure,  viz.
$\alpha$-ZnCl$_{2}$ ($I\bar{4}2d$, Z=4], $\beta$-ZnCl$_{2}$ ($P2_{1}/n$,
Z=12),  $\gamma$-ZnCl$_{2}$ ($P4_{2}/nmc$, Z=2), and $\delta$-ZnCl$_{2}$
($Pna2_{1}$, Z=4) in which the Zn atom is tetrahedrally coordinated with
Cl atoms. Secondly, ZnCl$_{2}$ displays unique  phase  transitions  upon
compression   as   well   as   pressure   release. 
For instance, the four-coordinated  $\alpha$-ZnCl$_{2}$ transforms into a 
six-coordinated CdCl$_{2}$-like phase at 2.1 GPa, which further
converts into four-coordinated $\gamma$-ZnCl$_{2}$ upon decompression\cite{saka}.  
Thirdly,   molten ZnCl$_{2}$, which is commercially used as a solvent and 
catalyst in  the hydro    cracking    of    coal    slurries    and   heavy  
crude   oil fractions\cite{gard}, typifies high viscosity\cite{mcke,grub,wood},  low
ionic     diffusivities\cite{bock},     and     very     low    electric
conductivity\cite{duke} due mainly to its non-uniform polymeric  network
linkage.\cite{ribe1}  Further,  liquid  ZnCl$_{2}$  has  the  ability to
supercool into a glass\cite{smith} at a relatively low glass  transition
temperature  (T$_{g}$  =  375  K).  Finally,  ZnCl$_{2}$ glass, which is
intermediate between strong (e.g.  SiO$_{2}$)  and  fragile  (e.g.  CKN)
glasses,  finds  its  potential  use  as a low-loss optical transmission
medium in the near infrared region.\cite{bryan}

\vspace{0.2cm}
\parindent 0.8cm
All  the crystalline phases of ZnCl$_{2}$ at ambient pressure are formed
by the network of ZnCl$_{4}$ tetrahedra sharing their corners. But  this
network  is  rather  soft (unlike SiO$_{2}$ or BeF$_{2}$) resulting in a
strongly   hybridized   character    for    the    normal    modes    of
vibration.\cite{smith}  Further,  in  the  low  frequency  range ($<$ 20
cm$^{-1}$), where the collective acoustic modes are expected to dominate
the  vibrational  density  of  states,  liquid  ZnCl$_{2}$  exhibits   a
prominent   boson  peak\cite{lebon,kart}  reflecting  its  more  complex
microstructure.  Ribero  et  al.\cite{ribe2}  examined  the  nature   of
instantaneous  normal  modes  in the atomic motion of liquid ZnCl$_{2}$,
which  involve  the  coupled  vibration   of   several   quasi-molecular
ZnCl$_{4}$ units of the network and has the local torsional character.
\vspace{0.2cm}
\parindent 0.8cm
Although  molecular  dynamics (MD) and Monte Carlo (MC) simulations have
been reported for glassy as well as molten ZnCl$_{2}$, the  need  for  a
realistic interatomic potential has remained unfulfilled.\cite{yann} The
present  work is primarily intended to make an attempt in formulating an
effective transferable interatomic potential that can be applied to  any
of   the   polymorphs   of   ZnCl$_{2}$   with  equal  ease.  Angell  et
al.\cite{ang1} had already suggested that there  is  little  distinction
between  the glassy and the crystalline behavior of ZnCl$_{2}$ so far as
the dynamics is concerned. They  further  opined  that  the  short  time
dynamics  of  liquid ZnCl$_{2}$ near its melting point (T$_{m}$ = 593 K)
might  be  regarded  as  solid-like  in  character.  Later   on,   X-ray
diffraction   measurements   of   molten   ZnCl$_{2}$   by   Triolo  and
Narten\cite{trio}  favored  this  argument  strongly.  Anomalously   low
enthalpy  and  volume  changes  on  fusion are supposed to be behind the
resemblance    of    ZnCl$_{2}$    melt     with     the     crystalline
structure.\cite{gard}  An extensive neutron diffraction study by Desa et
al.\cite{erw} further showed that vitreous ZnCl$_{2}$  has  a  distorted
closed  packed  array  of  Cl  atoms  in  which  Zn  atoms  are randomly
distributed in the tetrahedral sites. Numerous such studies  undoubtedly
imply  that  if  we  are  able to understand the dynamical properties of
crystalline polymorphs of  ZnCl$_{2}$,  microscopic  interpretation  for
glassy  and  molten  phases  will also be robust. In the present work we
have also determined the phonon density of states in the  $\alpha$-phase
by   coherent   inelastic   neutron   scattering.  The  potential  model
calculation is found to be in fair agreement with the neutron data.
                                                                                                                             
\section{Crystal structure}
Till  1961,  it  was  commonly believed that the ambient stable phase of
ZnCl$_{2}$ had a layered structure in which each Zn atom was coordinated
to six Cl atoms to form regular octahedra with  van  der  Waals  bonding
between   the   layers.\cite{paul}   Later  investigations  proved  that
ZnCl$_{2}$ has bonding based on a tetrahedral framework. Initially three
solid phases of ZnCl$_{2}$ were known crystallographically.  These  were
$\alpha$,  $\beta$  and  $\gamma$  polymorphs of ZnCl$_{2}$ belonging to
monoclinic and  tetragonal  crystal  classes  (Table~\ref{tab1}).  While
$\alpha$  and  $\beta$  phases  have cristobalite type three dimensional
networks of  corner  linked  tetrahedra\cite{breh},  $\gamma$-ZnCl$_{2}$
crystallizes  to  a  layered structure of cross linking tetrahedra as in
case  of  red  HgI$_{2}$.\cite{osw}  However,  in  1978,  Brynestad  and
Yakel\cite{bryn,yak} pointed out that all three modifications are rather
the  result  of water contamination and that purely anhydrous ZnCl$_{2}$
has an  orthorhombic  unit  cell  containing  corner  linked  ZnCl$_{4}$
tetrahedra.   This   orthorhombic   phase   has   been  referred  to  as
$\delta$-ZnCl$_{2}$ in  various  papers.\cite{saka,erw}  Fig.~\ref{fig1}
suggests  that  the  close  packed  layers  in  $\delta$-ZnCl$_{2}$  are
horizontal. Further, temperature has a significant impact\cite{ang2}  on
the  structure  of  ZnCl$_{2}$.  While  rapid  cooling  of  the melt (or
devitrification of the glass) produces the $\alpha$-phase, slow  cooling
of  the  melt  favors  the formation of $\beta$-ZnCl$_{2}$. On the other
hand,   concentrated   solution   yields   the   $\gamma$-phase   upon
crystallization.

\section{Experiments and data analysis}
Experimental   studies   involving   crystalline   ZnCl$_{2}$  encounter
difficulties because of its extreme hygroscopicity.  The  reagent  grade
chemical  ZnCl$_{2}$  (purity  98\%),  purchased from Merck Chemical Co.
(Germany), was dried for a period of 8 hours. It was then  exposed  into
the glove box under dry argon atmosphere for transferring into different
containers.  About  30  g  of  the  polycrystalline ZnCl$_{2}$ sample so
obtained was placed in a thin walled aluminum sample holder of  circular
shape  for  the  inelastic  neutron  scattering  (INS)  studies. Another
cylindrical vanadium  can  was  made  use  of  for  taking  the  neutron
diffraction  pattern  of  the  same  sample.  Neutron powder diffraction
confirmed that the sample belongs to the $\alpha$-phase. The  structural
refinements  of  the  powder  data  were  carried out using the Rietveld
profile refinement method with the program  DBWS  9411.\cite{young}  The
cell  constants and atomic coordinates refined from the data recorded at
300 K are in close agreement with the reported data.\cite{breh}
                                                                                                                             
\vspace{0.2cm}
\parindent 0.8cm
Inelastic  neutron  scattering  measurements  were  carried out at 300 K
using a medium resolution triple-axis  spectrometer\cite{chap3}  at  the
Dhruva  Reactor,  Trombay.  The  instrument used a monochromated beam of
neutrons produced by diffraction from  the  (111)  planes  of  a  copper
crystal.  Neutrons  scattered  inelastically  by  the  sample were again
diffracted by a  pyrolytic  graphite  (002)  analyzer  and  subsequently
detected  in  a $^{10}BF_{3}$ proportional counter. All the measurements
were made in the energy loss mode with constant momentum transfer ($\bf{Q}$).
The elastic energy resolution was about  15\%  of  the  initial  energy.
Several scans were performed with the final energy
($E_{f}$) values of 30 meV and $\bf{Q}$ values of
5 to 6 \AA$^{-1}$ that are much larger than the size of the Brillouin zone (about 1 \AA$^{-1}$). Additional
measurements of the neutron background were made by
detuning the analyzer crystal from the position by
$\pm 5^{o}$ and repeating the scan for the same
sample.

\vspace{0.2cm}
\parindent 0.8cm
 The neutron weighted phonon density of states, $g^{n}(E)$, are obtained
from   the   measured  scattering  function  $S(\bf{Q},E)$  through  the
following relation\cite{price}
\begin{eqnarray}
g^{n}(E) = A \left\langle {e^{2W(Q)} \over Q^2} {E\over n(E,T) + 1} S(\bf{Q}, E) \right\rangle \\
\approx B \sum_{p} {{4\pi b_{p}^2}\over M_{p}}g_{p}(E)
\label{eq5}
\end{eqnarray}
where  $n(E,T)$  =  $[exp(E/KT)-1]^{-1}$;  $A$,  $B$  are  normalization
constants; $2W(Q)$ is the Debye-Waller  factor;  $b_{p}$,  $M_{p}$,  and
$g_{p}(E)$ refer respectively to the neutron scattering length, mass and
partial  phonon density of states of the $p^{th}$ atom in the unit cell.
The  quantity  within  $\left\langle...\right\rangle$   represents   the
average  over  all $\bf{Q}$  values.  The factor ${4\pi b_{p}^2}\over M_{p}$
turns out to be  0.063  and  0.474  barn/a.m.u.  for  Zn  and  Cl  atoms
respectively.

\section{Model interatomic potential}
As previously stated, there  is  a  need  to  formulate a
realistic potential model that can effectively model the structure
and dynamics in various polymorphs of ZnCl$_{2}$.  In the present
model, Born-Mayer repulsive  and attractive van der Waals
interactions among different   pairs   of   atoms   have been
incorporated. Because of  the  covalent  nature  of  the Zn-Cl
bond, a stretching term has to be included further in the
potential model. The model thus turns  out  to  be
\begin{equation}
V(r)  =  A\hspace{0.1cm}  exp\left[ -\frac{B\hspace{0.1cm}r}{R(k)+R(k')}
\right]    -    \frac{C_{kk'}}{r^6}    -    D\hspace{0.1cm}     exp\left[
-{n{(r-r_{0})^2}\over 2r} \right] \label{eq1}
\end{equation}
where  $A$  (=1822  eV)  and  $B$  (=12.364)  are  two  constants,  used
extensively in several previous  works.\cite{chap1,mala,mittal};  $R(k)$
and  $R(k')$  refer  to effective radius parameters for Zn and Cl atoms.
$C_{kk'}$ accounts for van der Waals  terms  between  Zn-Zn,  Zn-Cl  and
Cl-Cl  atoms;  $r_{0}$ is the equilibrium Zn-Cl bond length; $D$ and $n$
are two adjustable parameters. 
At the observed lattice constants and atomic
positions in the unit cell, the parameters were so adjusted as to give nearly 
a zero internal stress and net force on each individual atom. Once the structural constraints are
met, the potential parameters were further adjusted
so that the calculated eigen frequencies are real
for all the wave vectors in the entire Brillouin zone. The
group theoretical analysis provides the symmetry
vectors necessary for block diagonalization of the
dynamical matrix. Table~\ref{tab2} lists the final values
of the optimized parameters. All the necessary calculations were made
using DISPR\cite{chap2} developed at Trombay. Lattice parameters and atomic
coordinates obtained as a result of the potential minimization for various polymorphs 
of ZnCl$_{2}$ are displayed in Tables{~\ref{tab3a},~\ref{tab3b},~\ref{tab3c} \& ~\ref{tab3d}}.
At high pressures, the equilibrium crystal structures were obtained
from the free energy minimization.

\section{Vibrational modes: Raman and infrared active}
A  group  theoretical classification of various zone-center phonon modes
for each of the crystalline modifications  in  ZnCl$_{2}$  is  given  in
Table~\ref{tab1}.  Experimental  mode assignments are available for only
two   polymorphs   of   ZnCl$_{2}$   (viz.    $\alpha$-ZnCl$_{2}$    and
$\gamma$-ZnCl$_{2}$).          Measured          optical          phonon
data\cite{saka,yann,ang2,james} are found to be in good  agreement  with
our calculated results (Tables{~\ref{tab4a},~\ref{tab4b},~\ref{tab4c} \&
~\ref{tab4d}}).  It  may  be noted that the lowest frequency mode in the
$\gamma$-phase (36 cm$^{-1}$) is almost half the frequency  of  that  in
the  $\alpha$-phase  (80 cm$^{-1}$). Further the pressure dependence of
Raman modes in $\alpha$, $\gamma$ and $\delta$ forms of  ZnCl$_{2}$  are
calculated.  Except  for  the  $\alpha$-phase,  in  all the others, the
calculated  pressure  dependence  compares  well  with   the   available
measured\cite{saka,pols}     data     (see    Fig.~\ref{fig2}).    Since
$\beta$-ZnCl$_{2}$ has as many as 108 phonon modes, pressure  dependence
of these modes is not shown here.

\section{Phonon dispersion relations and phonon density of states in $\alpha$-ZnCl$_{2}$ }
                                                                                                                             
As  Table~\ref{tab1} suggests, there is a total of 18 degrees of
freedom in $\alpha$-ZnCl$_{2}$, giving rise to 18 phonon
frequencies. Following are the representations of normal modes
along the three high symmetry directions (viz. $\Sigma$, $\Lambda$
and $\Delta$).
\begin{center}
$\Sigma(100)$: 8$\Sigma_{1}$ + 10$\Sigma_{2}$\\
$\Lambda(001)$: 4$\Lambda_{1}$ + 4$\Lambda_{2}$ + 5$\Lambda_{3}$ ($\Lambda_{3}$ being doubly degenerate)\\
$\Delta(110)$: 9$\Delta_{1}$ + 9$\Delta_{2}$\\
\end{center}
Fig.~\ref{fig3}  displays  the calculated phonon dispersion relations in
$\alpha$-ZnCl$_{2}$. Considerable dispersion and anti-crossing are found
for the branch in all  the  three  symmetry  directions  which  indicate
strong  hybridization.  There are no indications of any phonon softening
in the calculations in any of the four polymorphs at ambient pressure.
                                                                                                                             
\vspace{0.2cm}
\parindent 0.8cm
We have plotted  the measured phonon density of states along with
the  calculated  spectrum in Fig.~\ref{fig4}. The  experimental  one-phonon  spectra  are
obtained   by   subtracting   the   multiphonon  contribution  from  the
experimental data. The multiphonon contribution is  obtained  using  the
Sjolander  formalism.\cite{sjol} The experimental data (Fig.~\ref{fig4})
confirm two broad structures with a band gap around 25 meV. The measured
neutron data\cite{gal} for vitreous ZnCl$_{2}$ and the Raman density  of
states  data\cite{alio}  for molten ZnCl$_{2}$ are qualitatively similar
to the present neutron data on crystalline $\alpha$-ZnCl$_{2}$.
\vspace{0.2cm}
\parindent 0.8cm
The computed partial phonon density of states is shown in Fig 5(a). This
indicates that the Zn ions contribute significantly at energies below 10 meV
while  from  10  meV  to  19  meV,  the  vibrations of Cl ions dominate.
However, in the higher energy region, contributions from both  the  ions
remain  more  or  less  similar.  A  comparison of the calculated phonon
density of states for all the crystalline polymorphs is  shown  in  Fig.
5(b).  Two  band  gaps  (a broad one around 23 meV and a very narrow one
around 36 meV) are observed in each of the phases,  though  of  slightly
varying  degree.  For  these  calculations,  all  the  phonon  modes are
integrated at each wave vector within the irreducible Brillouin zone  on
a $10\times10\times10$ mesh with an energy resolution of 1.0 meV.
                                                                                                                             
\section{Thermodynamic properties}
                                                                                                                             
\subsection{Heat capacity and Debye temperature}
As Fig.~\ref{fig6}(a) suggests, there is a fairly good agreement
between the calculated and the experimental data\cite{ang3} on the molar heat capacity.
As the inset shows, there is vary small variation among heat capacities of various polymorphs, which reflects 
the small variation in the respective phonon density of states. In particular, the spike in $\Delta C_{p}$ at about 
50 K in the inset of Fig.~\ref{fig6}(a) arises due to the different phonon density of states of the $\alpha$-phase
at low energy in Fig.~\ref{fig5}(b). Fig.~\ref{fig6}(b) displays the Debye temperature ($\theta_{D}$) plot
for different crystalline modifications in ZnCl$_{2}$. The characteristic Debye temperature in the present model turns out to be 375 K, while the experimental value is 340 K\cite{ang3}.

\subsection{Mode  Gr\"uneisen parameters and thermal expansion}
The   mode  Gr\"uneisen  parameters  are  important  indicators  of  the
anharmonicity of phonon modes and their role in determining the  thermal
expansion. For the i$^{th}$ mode the Gr\"uneisen parameter is given by
                                                                                                                             
\begin{equation}
\gamma_{i} = - {\partial ln~\omega_{i}\over\partial ln~V}
\label{eq3}
\end{equation}
                                                                                                                             
In the quasiharmonic approximation, the ${\it i}$th phonon mode contributes ($1
\over   BV$)$\gamma_{i}$C$_{Vi}$   to   the   volume  thermal  expansion
($\alpha_{V}$), where $C_{Vi}$ denotes the contribution of the  $i^{th}$
mode  to  the  volume  heat  capacity. As heat capacity and bulk modulus
values remain similar for each of the polymorphs, it  is  obviously  the
mode  Gr\"uneisen  parameter  which  dominates  in  the determination of
$\alpha_{V}$.  Fig.~\ref{fig7}  displays  averaged  $\gamma_{i}$  as   a
function  of  phonon  energy  for  different  crystalline  polymorphs in
ZnCl$_{2}$. One may notice that while for $\alpha$, $\beta$ and $\delta$
phases,  18  meV  phonons   have   maximum   variations   with   volume,
$\gamma$-ZnCl$_{2}$   behaves  slightly  differently.  This  is  further
reflected  in  the  subsequent  thermal   expansion   calculation   (see
Fig.~\ref{fig7}).
                                                                                                                            
\subsection{Equation of state}
The  crystal  structure  parameters as a function of pressure (at T=0 K)
are calculated corresponding  to  the  minimum  of  Gibbs  free  energy.
Fig.~\ref{fig8}  shows  how  the  cell dimensions belonging to different
crystalline polymorphs of ZnCl$_{2}$ vary with pressure. It is  observed
that the compressibility is anisotropic. While $\bf{c}$ axis is the most
compressible  in  $\beta$  and  $\gamma$  forms of ZnCl$_{2}$, it is the
least  compressible  in  $\alpha$-ZnCl$_{2}$.  In   $\delta$-ZnCl$_{2}$,
$\bf{b}$  axis is found to be the most compressible. However, the volume
derivative with pressure in all the phases looks a bit alike,  as  their
bulk moduli are similar.
                                                                                                                             
\section{Conclusions}
We  have  developed  an  interatomic potential model for ZnCl$_{2}$ that
mimicks the vibrational and thermodynamic properties of  the  polymorphs
of  crystalline  ZnCl$_{2}$  and  reproduces fairly  well the inelastic
neutron scattering  data  for  the  $\alpha$-phase. The fact that the $\gamma$-phase
shows differences in Gr\"uneisen parameters and
lattice expansion is essentially due to its layered structure, which is
very different from other corner-shared polymorphs. This  transferable
potential  model  may be exploited further to investigate the properties
of the glassy and the liquid phases of ZnCl$_{2}$. 

\ack
A.  S.  expresses  his  gratitude  to  the  Council  of  Scientific  and
Industrial  Research  (CSIR,  New  Delhi),  India,  for  rendering   the
financial assistance and acknowledges as well the encouragement and care
taken  by  Dr.  M.  Ramanadham  and  Dr. V. C. Sahni. We thank Dr. S. N.
Achary for his kind cooperation in making the commercial polycrystalline
ZnCl$_{2}$  sample  suitable  for  the  inelastic   neutron   scattering
experiments. We are also thankful to the referees for their valuable suggestions.

\Figures
\begin{figure}
\caption{\label{fig1}Crystal
structures  of polymorphs    of    ZnCl$_{2}$,    viz.    (a)
$\alpha$-ZnCl$_{2}$,    (b) $\beta$-ZnCl$_{2}$, (c)
$\gamma$-ZnCl$_{2}$ and (d) $\delta$-ZnCl$_{2}$. The  small
circles represent Cl atoms, while Zn atoms are located inside (not
visible here) the ZnCl$_{4}$ tetrahedra.}

\caption{\label{fig2}Pressure dependence of phonon modes in (a)
$\alpha$-ZnCl$_{2}$     (b) $\gamma$-ZnCl$_{2}$ and    (c)
$\delta$-ZnCl$_{2}$.} 

\caption{\label{fig3}Calculated phonon
dispersion relations in $\alpha$-ZnCl$_{2}$.}

\caption{\label{fig4}Plots of (a) experimental and (b) calculated
neutron weighted  phonon density of states for
$\alpha$-ZnCl$_{2}$. The multiphonon contribution at 300 K has
been subtracted from the experimental data  to obtain the
experimental one-phonon spectrum. The corresponding calculated
spectrum has been convoluted with the energy resolution of the
instrument.} 

\caption{\label{fig5} (a) Calculated partial phonon
density of states of various atoms in $\alpha$-ZnCl$_{2}$. (b) A
comparison of calculated one phonon density of states in various
crystalline polymorphs of ZnCl$_{2}$.} 

\caption{\label{fig6}Plots
of  (a) heat capacity ($C_{P}$) and (b) Debye temperature ($\theta_{D}$)
for all the crystalline polymorphs of ZnCl$_{2}$. The
inset in (a) shows the specific heat differences of
various polymorphs with respect to the $\alpha$-phase.}

\caption{\label{fig7}Calculated  plots of  the mode Gr\"uneisen
parameter and the volume thermal expansion coefficients of
ZnCl$_{2}$  in  its various crystalline phases.}

\caption{\label{fig8}Calculated   equations  of  state for
crystalline ZnCl$_{2}$ in (a) $\alpha$ (b) $\beta$ (c) $\gamma$
and (d) $\delta$ phases.}
\end{figure}

\Tables
\Table{\label{tab1}Group theoretical classification of zone-centre phonon modes in various crystalline polymorphs of ZnCl$_{2}$. Raman and infra-red active phonon branches are indicated within the bracket by {\it{R}} and {\it{ir}} respectively.}
\br
\multicolumn{1}{c}{Phase}&\multicolumn{1}{c}{Crystal system}&\multicolumn{1}{c}{Space-group}&\multicolumn{1}{c}{Z}&\multicolumn{1}{c}{Group theoretical classification at zone-centre} \\ 
\mr
$\alpha$&Tetragonal&$I\bar{4}2d$&4&${A_{1}(R) + 2A_{2} + 2B_{1}(R) + 3B_{2}(R,ir) + 5E(R,ir)}$ \\
$\beta$&Monoclinic&$P2_{1}/n$&12&${27A_{g}(R) + 27A_{u}(ir) + 27B_{g}(R) + 27B_{u}(ir)}$ \\
$\gamma$&Tetragonal&$P4_{2}/nmc$&2&${A_{1g}(R) + 2A_{2u}(ir) + 2B_{1g}(R) + B_{2u}(ir) + 3E_{g}(R) + 3E_{u}(ir)}$ \\
$\delta$&Orthorhombic&$Pna2_{1}$&4&${9A_{1}(R,ir) + 9A_{2}(R) + 9B_{1}(R,ir) + 9B_{2}(R,ir)}$ \\
\br
\end{tabular}
\end{indented}
\end{table}

\Table{\label{tab2}Transferable potential parameters for ZnCl$_{2}$ polymorphs.}
\br
\multicolumn{2}{c}{$R$(\AA)}&\multicolumn{3}{c}{C$_{ij}$(eV\AA$^{6}$)}& $D(eV)$ & $n$(\AA$^{-1}$) & $r_{0}$(\AA)\\
Zn&Cl&Zn-Zn&Zn-Cl&Cl-Cl\\
\mr
1.03&2.52&20&135&550&3.00&4.60&2.30\\
\br
\end{tabular}
\end{indented}
\end{table}
                                                                                                          
\Table{\label{tab3a}Lattice constants and position parameters
for $\alpha$-ZnCl$_{2}$. For the space group $I\bar{4}2d$, Zn
atoms are located at (0,0,0) and $\left({1\over 2},0,{3\over 4}
\right)$ while Cl atoms are located at general positions of
$\left(x,{1\over 4},{1\over 8} \right)$, $\left(\bar{x},{3\over
4},{1\over 8} \right)$, $\left({1\over 4},\bar{x},{7\over 8}
\right)$, $\left({3\over 4},x,{7\over 8} \right)$ respectively.}
\br
&\multicolumn{1}{c}{This work}&\multicolumn{1}{c}{Experimental data} \\
&&\multicolumn{1}{c}{(Ref. 1)} \\  
\mr
{$a$(\AA)}&5.406&5.410 \\
{$c$(\AA)}&10.415&10.300 \\
V{(\AA$^{3}$)}&304.34&301.46 \\
\emph{x}&0.24&0.25  \\
\br
\end{tabular}
\end{indented}
\end{table}

\Table{\label{tab3b}Lattice constants and position parameters
for $\beta$-ZnCl$_{2}$. For the space group $P2_{1}/n$, Zn and Cl
atoms are located at general positions of $(x,y,z)$,
$\left(\bar{x}+{1\over 2},y+{1\over 2},\bar{z}+{1\over 2}
\right)$, $(\bar{x},\bar{y},\bar{z})$, $\left(x+{1\over
2},\bar{y}+{1\over 2},z+{1\over 2} \right)$ respectively.}
\br
&\multicolumn{3}{c}{This work}&\multicolumn{3}{c}{Experimental data} \\
&\multicolumn{3}{c}{}&\multicolumn{3}{c}{(Ref. 1)} \\  
\mr
$a$(\AA)&\multicolumn{3}{c}{6.519}&\multicolumn{3}{c}{6.500} \\
$b$(\AA)&\multicolumn{3}{c}{11.330}&\multicolumn{3}{c}{11.300} \\
$c$(\AA)&\multicolumn{3}{c}{12.272}&\multicolumn{3}{c}{12.300} \\
$\beta$($^{o}$)&\multicolumn{3}{c}{89.8}&\multicolumn{3}{c}{90.0}\\
V(\AA$^{3}$) &\multicolumn{3}{c}{906.43}&\multicolumn{3}{c}{903.43}\\
\mr
Atom&\emph{x}&\emph{y}&\emph{z}&\emph{x}&\emph{y}&\emph{z} \\
\mr
Zn&0.165&0.167&0.065&0.167&0.167&0.063 \\
Zn&0.166&0.499&0.186&0.167&0.500&0.188 \\
Zn&0.666&0.667&0.187&0.667&0.667&0.188 \\
Cl&0.328&0.003&0.127&0.333&0.000&0.125  \\
Cl&0.326&0.334&0.124&0.333&0.333&0.125  \\
Cl&0.335&0.663&0.128&0.333&0.667&0.125  \\
Cl&0.836&0.167&0.127&0.833&0.167&0.125  \\
Cl&0.835&0.503&0.127&0.833&0.500&0.125  \\
Cl&0.828&0.831&0.122&0.833&0.833&0.125  \\
\br
\end{tabular}
\end{indented}
\end{table}

\Table{\label{tab3c}Lattice  constants and position parameters
for $\gamma$-ZnCl$_{2}$. For the space  group $P4_{2}/nmc$,  Zn
atoms  are located  at (0,0,0) and $\left({1\over 2},{1\over
2},{1\over 2} \right)$ while  Cl  atoms are  located  at  general
positions of $\left(0,{1\over   2},z \right)$, $\left(0,{1\over
2},z+{1\over     2}     \right)$, $\left({1\over
2},0,\bar{z}+{1\over 2} \right)$, $\left({1\over 2},0,\bar{z}
\right)$ respectively.}
\br
& {This work} & {Experimental data} \\
& & {(Ref. 1)} \\  
\mr
$a$(\AA)& {3.733} & {3.700} \\
$c$(\AA)& {10.859} & {10.670} \\
V(\AA$^{3}$) & {151.31} & {146.07} \\
\emph{z} & 0.121& 0.125  \\
\br
\end{tabular}
\end{indented}
\end{table}

\Table{\label{tab3d}Lattice constants and position parameters
for $\delta$-ZnCl$_{2}$. For the space group $Pna2_{1}$, Zn and Cl
atoms are located at general positions of $(x,y,z)$,
$\left(\bar{x},\bar{y},z+{1\over 2} \right)$, $\left(x+{1\over
2},\bar{y}+{1\over 2},z \right)$, $\left(\bar{x}+{1\over
2},y+{1\over 2},z+{1\over 2} \right)$ respectively. U(\AA$^{2}$)
gives the isotropic thermal amplitudes.}
\br
&\multicolumn{4}{c}{This work} & \multicolumn{4}{c}{Experimental data} \\
&\multicolumn{4}{c}{} & \multicolumn{4}{c}{(Ref. 2)} \\  
\mr
$a$(\AA)&\multicolumn{4}{c}{6.444} & \multicolumn{4}{c}{6.443} \\
$b$(\AA)&\multicolumn{4}{c}{7.666} & \multicolumn{4}{c}{7.693} \\
$c$(\AA)&\multicolumn{4}{c}{6.122} & \multicolumn{4}{c}{6.125} \\
$\beta$($^{o}$)&\multicolumn{4}{c}{90.0} & \multicolumn{4}{c}{90.0}\\
V(\AA$^{3}$)&\multicolumn{4}{c}{302.40} &
\multicolumn{4}{c}{303.59}\\ 
\mr
Atom&\emph{x}&\emph{y}&\emph{z}&U&\emph{x}&\emph{y}&\emph{z}&U \\
\mr
Zn&0.081&0.125&0.378&0.0303&0.0818&0.1251&0.3750&0.0253 \\
Cl&0.075&0.122&0.005&0.0296&0.0702&0.1223&0.0041&0.0305  \\
Cl&0.084&0.630&0.003&0.0296&0.0841&0.6332&0.0062&0.0292  \\
\br
\end{tabular}
\end{indented}
\end{table}

\Table{\label{tab4a}Comparison of calculation and experimental
data of optical modes in $\alpha$-ZnCl$_{2}$. Modes of $A_{2}$
representation are not optically active.}
\br
\multicolumn{7}{c}{Optical phonon modes ($cm^{-1}$)} \\
\multicolumn{1}{c}{Representations} & \multicolumn{1}{c}{This work} & \multicolumn{5}{c}{Experimental data} \\
\multicolumn{1}{c}{} & \multicolumn{1}{c}{} &
\multicolumn{1}{c}{(Ref. 3)} & \multicolumn{1}{c}{(Ref. 23)} &
\multicolumn{1}{c}{(Ref. 31)} \\  
\mr
$A_{1}$&223&226&233&245 \\
$A_{2}$&147&&&  \\
&301&&&  \\
$B_{1}$&94&117&&113  \\
&273&&& \\
$B_{2}$&111&128&&128 \\
&325&&& \\
$E$&75&76&82&80 \\
&99&100&103&103 \\
&263&&& \\
&327&&& \\
\br
\end{tabular}
\end{indented}
\end{table}

\Table{\label{tab4b}Comparison of calculation and experimental
data of optical modes in $\beta$-ZnCl$_{2}$.}
\br
\multicolumn{8}{c}{Optical phonon modes ($cm^{-1}$)} \\
\multicolumn{4}{c}{This work} & \multicolumn{2}{c}{Experimental data} \\
\multicolumn{4}{c}{} & \multicolumn{1}{c}{(Ref. 31)} & \multicolumn{1}{c}{(Ref. 23)} \\
\multicolumn{1}{c}{$A_{g}$} & \multicolumn{1}{c}{$B_{g}$} &
\multicolumn{1}{c}{$A_{u}$} & \multicolumn{1}{c}{$B_{u}$} &
\multicolumn{1}{c}{($A_{g}$,$B_{g}$})&
\multicolumn{1}{c}{($A_{u}$,$B_{u}$})\\ 
\mr
34&33&34&36 \\
40&47&43&52&&42 \\
44&49&48&57&& 57 \\
47&55&53&61&&61 \\
48&61&58&69&&70 \\
57&66&64&74 \\
64&71&70&84 \\
70&75&93&86&78& \\
72&79&94&100&&99 \\
90&91&108&107&104&105 \\
102&104&110&116&108& \\
114&111&120&127&118&123 \\
122&114&120&144&125&135 \\
135&136&141&222 \\
148&138&216&232&231& \\
222&221&231&254&250 &\\
241&236&252&257 \\
248&248&261&266&268 & \\
261&260&264&273&278 &\\
266&267&270&308 \\
273&268&308&317 \\
311&312&316&324 \\
315&320&324&326 \\
324&322&327&328 \\
327&327&328&340 \\
328&328&340 \\
340&340 \\
\br
\end{tabular}
\end{indented}
\end{table}

\Table{\label{tab4c}Comparison of calculation and experimental
data of optical modes in  $\gamma$-ZnCl$_{2}$.}
\br
\multicolumn{4}{c}{Optical phonon modes ($cm^{-1}$)} \\
\multicolumn{1}{c}{Representations} & \multicolumn{1}{c}{This work} & \multicolumn{2}{c}{Experimental data} \\
\multicolumn{1}{c}{} & \multicolumn{1}{c}{} &
\multicolumn{1}{c}{(Ref. 3)} & \multicolumn{1}{c}{(Ref. 31)} \\
\mr
$A_{1g}$&244&248&252 \\
$A_{2u}$&258  \\
$B_{1g}$&66  \\
&271 \\
$B_{2u}$&216 \\
$E_{g}$&44&36&38 \\
&92&88&80 \\
&322 \\
$E_{u}$&68 \\
&318 \\
\br
\end{tabular}
\end{indented}
\end{table}

\Table{\label{tab4d}Comparison of calculation and experimental
data of optical modes in $\delta$-ZnCl$_{2}$.}
\br
\multicolumn{6}{c}{Optical phonon modes ($cm^{-1}$)} \\
\multicolumn{4}{c}{This work}&\multicolumn{2}{c}{Experimental data} \\
\multicolumn{4}{c}{}&\multicolumn{1}{c}{(Ref. 3)}&\multicolumn{1}{c}{(Ref. 16)} \\
\multicolumn{1}{c}{$A_{1}$}&\multicolumn{1}{c}{$A_{2}$}&\multicolumn{1}{c}{$B_{1}$}&\multicolumn{1}{c}{$B_{2}$} \\ 
\mr
45&50&50&72&&39,49,59 \\
75&55&67&91&&74,83,88 \\
92&76&108&118&&96,100,102,106 \\
112&104&128&141&147&109,124,129 \\
222&128&242&267&&226,250 \\
264&235&270&269&278&272,282 \\
324&269&317&298&&321 \\
329&318&326&329 \\
&327 \\
\br
\end{tabular}
\end{indented}
\end{table}

\end{document}